\documentstyle[12pt,epsfig]{article}
\def\beq{\begin{equation}}
\def\eeq{\end{equation}}
\def\bea{\begin{eqnarray}}
\def\eea{\end{eqnarray}}

\begin{document}

\begin{center}
{\Large \bf \sf Critical coupling and coherent perfect absorption for ranges of energies
due to a complex gain and loss symmetric system
  }

\vspace{1.3cm}

{\sf Mohammad Hasan \footnote{e-mail address: \ \ mohammadhasan786@gmail.com} Ananya Ghatak \footnote{e-mail address: \ \ gananya04@gmail.com}
and Bhabani Prasad Mandal \footnote{e-mail address:
\ \ bhabani.mandal@gmail.com, \ \ bhabani@bhu.ac.in  }}

\bigskip

{\em $^*$ISRO Satellite Centre (ISAC),
Bangalore-560017, INDIA \\
$^{\dagger \ddagger}$Department of Physics,
Banaras Hindu University,
Varanasi-221005, INDIA. \\ }

\bigskip
\bigskip

\noindent {\bf Abstract}

\end{center}

We consider a non-Hermitian medium with a gain and loss symmetric, exponentially damped potential distribution to demonstrate different 
scattering features analytically. The condition for critical coupling (CC) for unidirectional 
wave and coherent perfect absorption (CPA) for bidirectional waves are obtained analytically
for this system. The energy points at which total absorption occurs are shown to be the 
spectral singular points for the time reversed system. The possible 
energies at which CC occurs for left and right incidence are different. We further obtain periodic intervals 
with increasing periodicity of energy for CC and CPA to occur in this system.

\medskip
\vspace{1in}
\newpage

\section{Introduction}

The recent ideas of $PT$-symmetric non-Hermitian quantum mechanics \cite{ ben4,mos, ben} have 
been fruitfully extended to optics due to formal equivalence between 
Schroedinger equation and certain wave equations in optics \cite{ opt1, opt2, opt3}. The 
parity operator $P$ stands for
spatial reflections $(x\rightarrow -x, p\rightarrow -p)$, while the anti-liner time reversal operator $T$
leads to  $(i\rightarrow -i, p\rightarrow -p, x\rightarrow x)$. The
equivalence between quantum mechanics and optics becomes possible due to the 
judicious inclusion of complex refractive index distribution
$V(x)=\eta_R(x)+i\eta_I(x)$, in the electromagnetic wave equation \cite{opt2, eqv1}. 
To realize this consider a one dimensional optical system with effective refractive index $\eta_R(x)+i\eta_I(x)$
in the background of constant refractive index $\eta_0$, $\eta_I$ stands for gain and loss component. The electric field
$E=E_0(x,z)e^{i(wt-kz)}$ of a light wave propagating in this medium (with $\eta_0>>\eta_I,\eta_R$) satisfies
the Schroedinger like equation,
\begin{eqnarray}
i\frac{\partial}{\partial z}E_0(x,z)&=&\Big[\frac{1}{2k}\frac{\partial^2}{\partial x^2}+k_0[\eta_R(x)+i\eta_I(x)]\Big]E_0(x,z) \nonumber \\
&=& HE_0(x,z) 
\end{eqnarray}
$k=k_0\eta_0$, $k_0$ being the wave number in vacuum. This Hamiltonian $H$ has gain and loss symmetry for $\eta_R(x)=
\eta_R(-x)$ and $\eta_I(x)=-\eta_I(-x)$. Thus complex optical gain and loss potential can be realized by judiciously incorporating
gain and loss profile on an even index distribution.
This important realization opens a wide window to study optical systems with gain and loss medium and leads
to the experimental observation of PT-invariance and its breaking \cite{bm}-\cite{sol} in various optical systems 
\cite{opt1, opt2, opt3,eqv1,opap, opapp}. 

Various features of quantum
scattering due to non-Hermitian potentials like exceptional points \cite{ep0}-\cite{ep2} and 
spectral singularity (SS) \cite{ss1}-\cite{ss3}, reflectionlessness and invisibility 
 \cite{aop, inv2, inv1, inv3}, reciprocity  \cite{aop, ss3, resc} etc have generated huge 
interest due to their applicability and usefulness in the study of optics. Recently the 
observation of perfect absorption \cite{cpa00}-\cite{cc4} of
incident electromagnetic wave by an optical media with complex refractive index distribution  
is counted as a big achievement in optics. The coherent perfect absorber (CPA)
which is actually the time-reversed counterpart of a laser
has become the center to all such studies
in optics due to the discovery of anti-laser \cite{cpa00, cpa01, cpa011} in which incoming 
beams of light interfere with one another in such a way as to perfectly cancel each other out. 
This phenomena of perfect absorption in optics can also be observed in quantum scattering 
when particles (with a mass $m$ and total energy $E$) interact with the surrounding medium through a 
complex potential distribution $V(x)$. 

Scattering due to complex potential can be described in a simple mathematical language as follows.
If A and B are the incident wave amplitudes from left and right directions and
C and D are the outgoing wave amplitudes to left and right respectively, then the scattering amplitudes are related
through scattering matrix as,
\begin{eqnarray} 
\left(\begin {array}{clcr}
C  \\
D \\
\end{array} \right) &=& 
S
\left(\begin {array}{clcr}
A  \\
B \\
\end{array} \right) , \ \ \mbox{where} \ \ S=\left(\begin {array}{clcr}
t_l & r_r  \\
r_l & t_r \\
\end{array} \right) .
\label{s}
\end{eqnarray}
For the perfect absorption the outgoing amplitudes C and D vanish leading to,
\begin{eqnarray}
t_lA +r_rB=0 \ ; \ \ \ r_lA+t_rB=0 \ ;
\label{rtrt}
\end{eqnarray}
The condition for perfect absorption for unidirectional incident waves can be written as, 
\begin{eqnarray}
t_l=0 \ ; r_l&=&0 , \ \ \mbox{for left incidence} \ (B=0) \nonumber \\
t_r=0 \ ; r_r&=&0 , \ \ \mbox{for right incidence} \ (A=0)
\label{ccc}
\end{eqnarray}
These situations are known in literature as critical coupling (CC) \cite{cc0}-\cite{cc4}. On the other hand
for bidirectional incident wave the condition for perfect absorption is,
\begin{equation}
\Big|det[S]\Big|=\Big| t_lt_r -r_lr_r\Big|=0
\label{cpa}
\end{equation}
This condition refers as coherent perfect absorption which has recently created lots of 
excitements \cite{cpa00}-\cite{cpa4} due to the discovery of anti-laser. This could pave the
way for a number of novel technologies with various applications from optical computers to
radiology \cite{cpa00, cpa011}. Thus it is extremely important to investigate different 
aspects of CPA using different non-Hermitian systems.

The purpose of this work is to investigate various prospectives of scattering of particles 
due to a complex potential. In particular we would like to study different aspects of null 
scattering 
(CC and CPA) and super scattering in the context of a particular non-Hermitian gain and loss symmetric
system which shows rich scattering properties. We consider non-Hermitian PT-symmetric version
of an exponentially damped optical potential to derive the condition for 
CC and CPA analytically. This specific potential is rather known as Wood-Saxon (WS) \cite{wsu1,wsu2}
potential, plays an important role in describing microscopic particle interactions.
CPA never 
happens for Hermitian or PT-symmetric non-Hermitian systems as $\Big|det[S]\Big|$ is always unity in these
cases. Here the complex PT-symmetric Wood-Saxon potential with an additive imaginary shift to 
the real axis shows scattering spectrum with total absorption for discrete as well as continuous ranges of energies. We 
further 
explicitly show the energy points at which CC or CPA occur correspond to SS points of the 
time reversed system. So far discrete energy points for CC and CPA are obtained for different
optical media. We are able to obtain narrow but finite ranges of energies for CC and CPA to 
occur in this system. The nature of CC depends on the direction of incident wave. 
No energy range exists for right incidence. This happens due to the asymmetric 
left-right asymptotic behavior of this particular potential.
More interestingly these ranges occur periodically with increasing 
periodicity for this particular system. This demonstration can give ideas to build perfect absorbers
of matter waves which would be flexible to work for different ranges of energies of the incident particles.
As the potential distribution is analogous to the medium's refractive index 
profile this work can be extended to quantum optics for the absorption of electromagnetic waves. 

Now we present the plan of this paper. In section II we discuss the scattering of non-Hermitian PT-symmetric WS potential and its
time reversal partner potential. The CC for this system at discrete positive energies is
reported in Sec. III. In Sec. IV we calculate the periodic ranges for CC in this potential. 
Sec. V is devoted to obtain the condition for CPA and its ranges while Sec. VI is left for 
conclusions and discussions.

\section{Scattering from gain and loss symmetric WS potential}
In this section we explore the nature of scattering when the single-particle
space in expanded in a Wood-Saxon basis \cite{wsbs}. If particle scatters through a medium 
consists of semi-infinite nuclear matter it is more reasonable to take the nucleon distribution
of a Wood-Saxon type rather than an uniform distribution \cite{wsdf, wsds}.
The Hermitian WS potential in a simplified form \cite{ws2} can be written as ,
\begin{equation}
V(x)=-\frac{V_{0}}{1+e^{\frac{x-r_0}{a}}} \equiv -\frac{V_{0}}{1+qe^{\delta x}} \ ,
\label{ws}
\end{equation}
where $V_0$ is the WS potential depth, $r_0$ is the width of the potential i.e the nuclear 
radius and $a$ is the diffuseness of semi-infinite nuclear matter \cite{wsdf}. The WS-potential
is rewritten in a more simplified form in Eq. (\ref{ws}) with $\delta =1/a$ and 
$q=e^{-r_0\delta }$, for its quantum mechanical study.
In this simplified form $V_0$ and $\delta $ determines the height and shape of the potential 
respectively.
For positive $\delta $ and $V_0$ this 
potential asymptotically becomes $-V_0$ for $x\rightarrow 
-\infty$ and vanishes for $x\rightarrow +\infty$. The solutions of Schroedinger equation 
for this potential correspond to both bound states and scattering states and are useful in 
studying different problems. Two independent scattering state solutions of the particle wave equation (with $q=1$ and $\hbar=1$) for WS potential are 
written as \cite{ws2},  
\begin{eqnarray}
\psi_1(x)&=& M(1+e^{\delta  x})^{-\alpha _2-\alpha _3} \ e^{ \alpha _3\delta x}\nonumber \\
&\times & _2F_1(1+\alpha _2+\alpha  _3,\alpha _2+\alpha _3;1+2\alpha _2; \frac{1}{1+e^{\delta  x}}) \ ;
\label{psi}
\end{eqnarray} 
\begin{eqnarray}
\mbox{and} \ \ \ \ \psi_{2} (x)&=& M \Big(\psi_1(x)\mid_{\alpha _2\rightarrow -\alpha _2}\Big) \ ,
\ M=(-1)^{2\alpha  _3 }\frac{\Gamma(\alpha _2-\alpha _3)\Gamma(1+\alpha _2-\alpha _3)}{\Gamma(1+2\alpha _2))\Gamma(-2\alpha_3)} \ ,
\label{psi2}
\end{eqnarray}
\begin{equation}
\label{a2}
\mbox{with} \ \ \ \ \ \alpha _{2}=\frac{2i}{ \delta } \sqrt{m E} \equiv \frac{2i}{ \delta } k_1; \ \ \
\alpha  _{3}=\frac{2i}{ \delta  } \sqrt{m (E+V_{0})}  \equiv \frac{2i}{ \delta  } k_2 \ .
\end{equation}
One can calculate the different scattering amplitudes by considering the superposition
of these two independent scattering state solutions and looking at the asymptotic behaviors of 
these.
We complexify the WS potential by taking the shape parameter imaginary (i.e. $\delta 
\rightarrow i\rho  $) such that it becomes PT-symmetric. 
Moreover for latter convenience 
we would like to consider this complex potential with an imaginary shift to the real axis as
\cite{ims, imsh},
\begin{equation}
\tilde {V}(\bar{x})=-\frac{V_{0}}{1+\bar{q}e^{i \rho  \bar{x}}}, \ \mbox{where} \ \bar{x}=x-i \zeta \ , \ 
 \zeta \ \ \mbox{is real,}
\label{pt1}
\end{equation} 
where $\bar{q}=e^{-ir_0\rho}=1$ and the complex part of the potential $\tilde {V}(\bar{x})$ working as the gain and loss component for this complex system. \\

\includegraphics[scale=0.75]{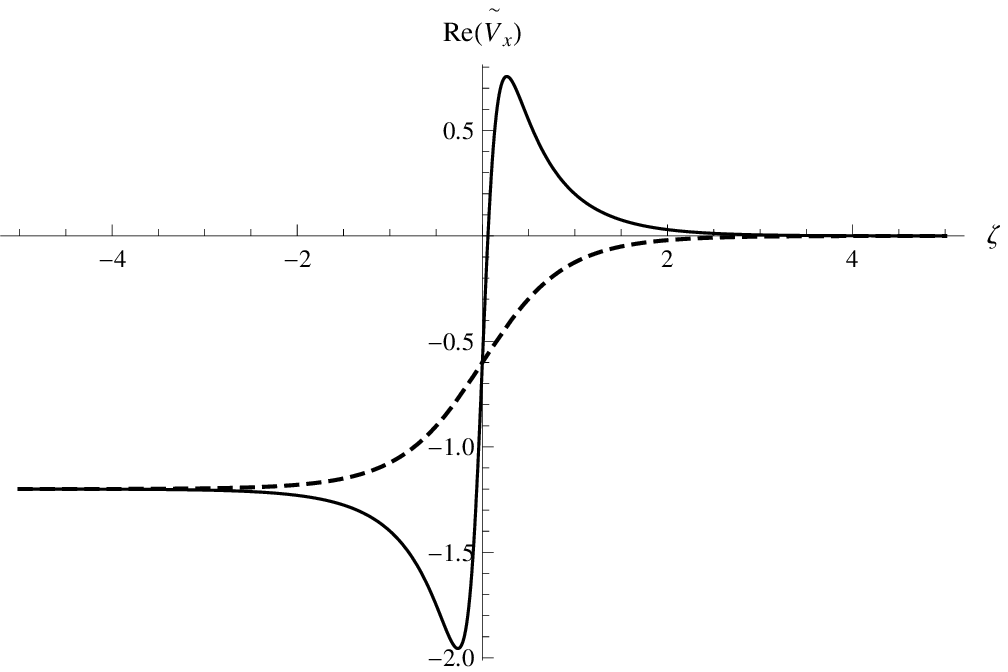} \ \ \ \includegraphics[scale=0.75]{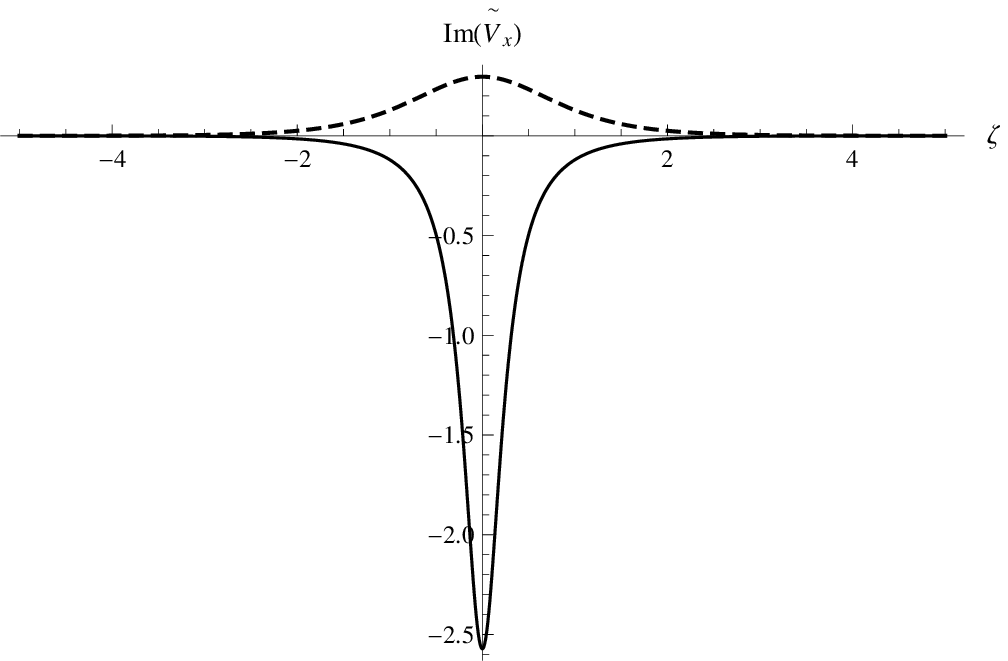} \\

{\it Fig. 1: Distributions of the real and imaginary components of PT-symmetric complex 
potential 
$\tilde {V}(\bar{x})$ are shown against $\zeta $ (for $V_0 = 1.2$ and $\rho  = 1.8$) 
with the values $x=2$ (continuous lines) and $x=4$ (dashed lines). } \\

The Schroedinger equation with respect to $\bar{x}$ will have the same form as with respect to real $x$. In this case $x$ bears the periodic 
nature of $\tilde {V}(\bar{x})$ where $\zeta $ decides the asymptotic behavior of the potential. The complexified WS potential in 
Eq. (\ref{pt1}) has the same asymptotic behaviors as $\{x, \zeta \}\rightarrow +\infty$ and
$\{x, \zeta \}\rightarrow -\infty$ shown in Fig. 1. This observation leads us to obtain the scattering state 
solutions of PT-symmetric non-Hermitian WS potential in the same form as in the Eq. (\ref
{psi}) and (\ref{psi2}) subjected to the following modifications,
\begin{eqnarray}
x\rightarrow \bar{x} \ ; \ \ \alpha _2\rightarrow a_2=\frac{2}{\rho } \sqrt{m E}\equiv \frac{2}{\rho } k_1 \ ; \nonumber \\
\alpha _3\rightarrow a_3=\frac{2}{\rho } \sqrt{m (E+V_0)}\equiv \frac{2}{\rho } k_2 \ . 
\end{eqnarray}
The above solutions for the non-Hermitian case have the following asymptotic 
behaviors, 
\begin{eqnarray}
\label{sol}
\psi_1(\bar{x}\rightarrow +\infty)&=& Me^{-ik_1x}.e^{-k_1\zeta }=Me^{-ik_1\bar{x}}; \nonumber  \\  \ \psi_1(\bar{x}\rightarrow -\infty)&=& M\Big[G1 \ e^{ik_2x}.e^{ k_2\zeta }+G2 \ e^{-ik_2x}.e^{-k_2\zeta }\Big]=M\Big[G1 \ e^{ik_2\bar
{x}}+G2 \ e^{-ik_2\bar{x}}\Big]; \nonumber  \\
\psi_{2}(\bar{x}\rightarrow +\infty)&=& Me^{ik_1x}.e^{k_1\zeta }=Me^{ik_1\bar{x}} ; \nonumber  \\
\psi_{2}(\bar{x}\rightarrow -\infty)&=& M\Big[G3 \ e^{ik_2x}.e^{ k_2\zeta }+G4 \ e^{-ik_2x}.e^{- k_2\zeta }\Big] 
=M\Big[G3 \ e^{ik_2\bar{x}}+G4 \ e^{-ik_2\bar{x}}\Big] \ ,\nonumber  \\
\end{eqnarray}
where G1, G2, G3, and G4 are given in terms of Gamma functions as,
\begin{eqnarray}
G1 &=& \frac{\Gamma(1 + 2 a_2) \Gamma(-2 a_3)}{\Gamma(a_2 - a_3) \Gamma(1 + a_2 - a_3)};\nonumber \\
G2 &=& \frac{\Gamma(1 + 2 a_2) \Gamma(2 a_3)}{\Gamma(a_2 + a_3) \Gamma(1 + a_2 + a_3)} ;\nonumber \\
G3 &=& \frac{\Gamma(1 - 2 a_2) \Gamma(-2 a_3)}{\Gamma(-a_2 - a_3) \Gamma(1 - a_2 - a_3)};\nonumber \\
G4 &=& \frac{\Gamma(1 - 2 a_2)\Gamma(2 a_3)}{\Gamma(-a_2 + a_3) \Gamma(1 - a_2 + a_3)} \ . 
\label{gg}
\end{eqnarray}
The scattering wavefunctions from the solutions in Eq. (\ref{sol}) asymptotically diverges as the potential becomes $0$ and
$-V_0$ for $\zeta\rightarrow \pm \infty$ in the Fig. 1. Different scattering amplitudes for this gain and loss symmetric complex WS potential thus can be 
readily read out as, 
\begin{equation}
r_{l} = \frac{G4}{G3}=\frac{\Gamma(2 a_3)\Gamma(-a_2 - a_3) \Gamma(1 - a_2 - a_3)}{\Gamma(-2 a_3)\Gamma(-a_2 + a_3) \Gamma(1 - a_2 + a_3)} ;
\label{rl}
\end{equation}
\begin{equation}
t_{l} = \sqrt{\frac{k_1}{k_2}} \ \frac{1}{G3}= \sqrt{\frac{k_1}{k_2}} \ \frac{\Gamma(-a_2 - a_3) \Gamma(1 - a_2 - a_3)}{\Gamma(1-2 a_2)\Gamma(-2 a_3)} =t_r;
\label{t}
\end{equation}
\begin{equation}
r_{r} = -\frac{G1}{G3}=-\frac{\Gamma(1+2 a_2)\Gamma(-a_2 - a_3) \Gamma(1 - a_2 - a_3)}{\Gamma(1-2 a_2)\Gamma(a_2 - a_3) \Gamma(1 + a_2 - a_3)} \ .
\label{rr}
\end{equation}
It is easy to see that $|r_l|^2\equiv R_{l}\not=R_{r}\equiv |r_r|^2$ as $ a _2$
$a_3$ are real, as expected for PT-symmetric non-Hermitian systems. However we have 
$|t_l|^2\equiv T_{l}=T_{r}\equiv |t_r|^2$ in this case unitarity is violated, $R+T\not=1$.
On the other hand reciprocity and unitarity are restored for Hermitian case as $G1^*=G4$ and 
$G2^*=G3$, since $ \alpha  _2$ and $\alpha _3$ are purely imaginary.
 
Under time reversal transformation the WS potential becomes,
\begin{equation}
\tilde {V}^*(\bar{x})=-\frac{V_{0}}{1+e^{-i \rho  \bar{x}^*}} \ .
\label{pt2}
\end{equation}
The scattering amplitudes for $\tilde {V}^*(\bar{x})$ (denoted with prime) are obtained from that of $\tilde 
{V}(\bar{x})$ by changing the parameter $\rho \rightarrow -\rho $ i.e. $a_2\rightarrow -a_2$ and $a_3\rightarrow -a_3$ as,
\begin{equation}
r'_{l} = r_{l}\Big|_{\stackrel{a_2\rightarrow -a_2} {a_3\rightarrow -a_3}}=\frac{\Gamma(-2 a_3)\Gamma(a_2 + a_3) \Gamma(1 + a_2 + a_3)}{\Gamma(2 a_3)\Gamma(a_2 - a_3) \Gamma(1 + a_2 - a_3)} \ ;
\label{rl2}
\end{equation}
\begin{equation}
t'_{l} = t_{l}\Big|_{\stackrel{a_2\rightarrow -a_2} {a_3\rightarrow -a_3}}= \sqrt{\frac{k_1}{k_2}} \ \frac{\Gamma(a_2 + a_3) \Gamma(1 + a_2 + a_3)}{\Gamma(1+2 a_2)\Gamma(2 a_3)} =t'_r \ ;
\label{t2}
\end{equation}
\begin{equation}
r'_{r} = r_{r}\Big|_{\stackrel{a_2\rightarrow -a_2} {a_3\rightarrow -a_3}}=-\frac{\Gamma(1-2 a_2)\Gamma(a_2 + a_3) \Gamma(1 +a_2 + a_3)}{\Gamma(1+2 a_2)\Gamma(a_2 + a_3) \Gamma(1 - a_2 + a_3)} \ .
\label{rr2}
\end{equation}

Different properties of null scattering and super scattering will be discussed using these scattering amplitudes in the following sections.\\

\section{Total absorption of unidirectional wave: Critical Coupling}

Particle waves of certain specific frequencies when incident from one direction on a potential are 
completely absorbed by the potential due to critical coupling (CC). In this section we discuss
the CC due to non-Hermitian gain and loss symmetric WS potential and calculate the frequencies which 
are absorbed by the system. The transmission and reflection coefficients become 
identically 
zero at these energies. We observe that the frequencies of the waves for CC for left incidence are different from that of for right incidence.

\subsection{CC for left incident wave}

The transmission and reflection coefficients of $\tilde{V}(\bar{x})$ are written from Eqs. (
\ref {rl}) and (\ref{t}) as, 
\begin{equation}
R_l=\mid r_l\mid^2=\mid\frac{G4}{G3}\mid^2 \ ;
T_l=\mid t_l\mid^2=\mid\frac{1}{G3}\mid^2 \ .
\end{equation}
$R_l$ and $T_l$ become simultaneously zero if $2a_3=n$, a positive integer. This happens
for the positive discrete energy,
\begin{equation}
\label{cce}
E_{n}^l= \frac{\rho ^{2}}{16m} n^{2}-V_{0}, \ \mbox{with} \ \ n> \frac{4}{\rho }\sqrt{mV_0} \ .
\label{e1}
\end{equation}
This physically means when matter wave with energy $E_n^l$ is incident on a
medium with potential distribution $\tilde{V}(\bar{x})$
then the wave will be completely absorbed as $R_l=0$ and $T_l=0$. 
The successive energy separations for this null scattering,
\begin{equation}
\Delta E_n^l\equiv E_{n+1}^l-E_{n}^l=\frac{\rho  ^{2}}{16m} (2n+1)
\label{es}
\end{equation} 
are independent of depth of the potential. Now we consider the time reversed case of  
$\tilde{V}(\bar{x})$ with reflection and transmission amplitudes $r'_{l,r}$ and $t'_{l,r}$
 as given in Eqs. (\ref
{rl2}),(\ref
{rr2}) and (\ref{t2}). In this case $R'_l$ diverges when $2a_3=n$, leading to the SS at the same 
energy point $E_n^l$. Thus we analytically see CC of a gain and loss symmetric non-Hermitian WS
potential are the SS points of the time reversed of the same potential. This is demonstrated 
graphically in Fig. 2.  \\

\includegraphics[scale=0.8]{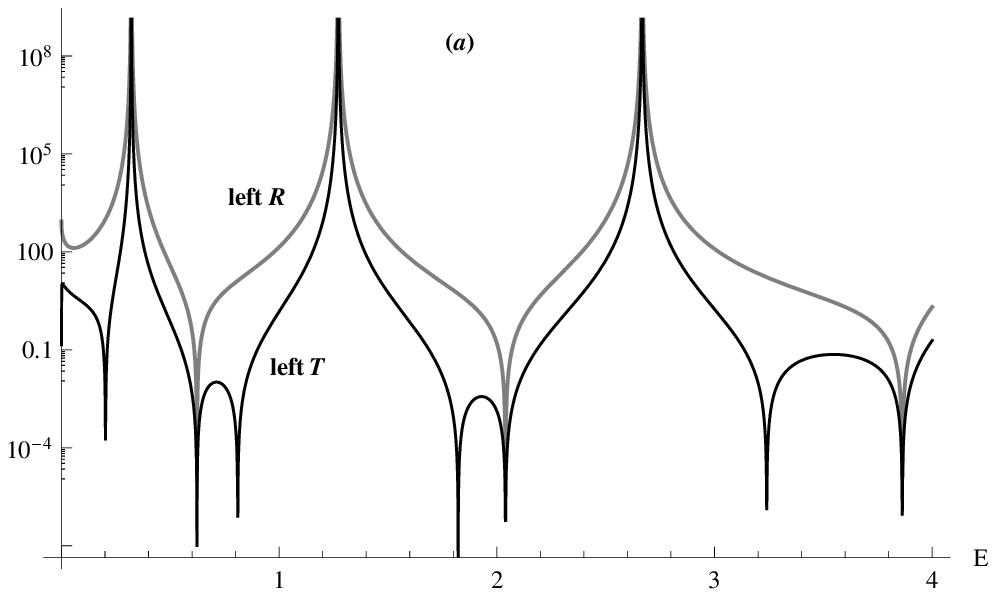} \ \ \ \includegraphics[scale=0.78]{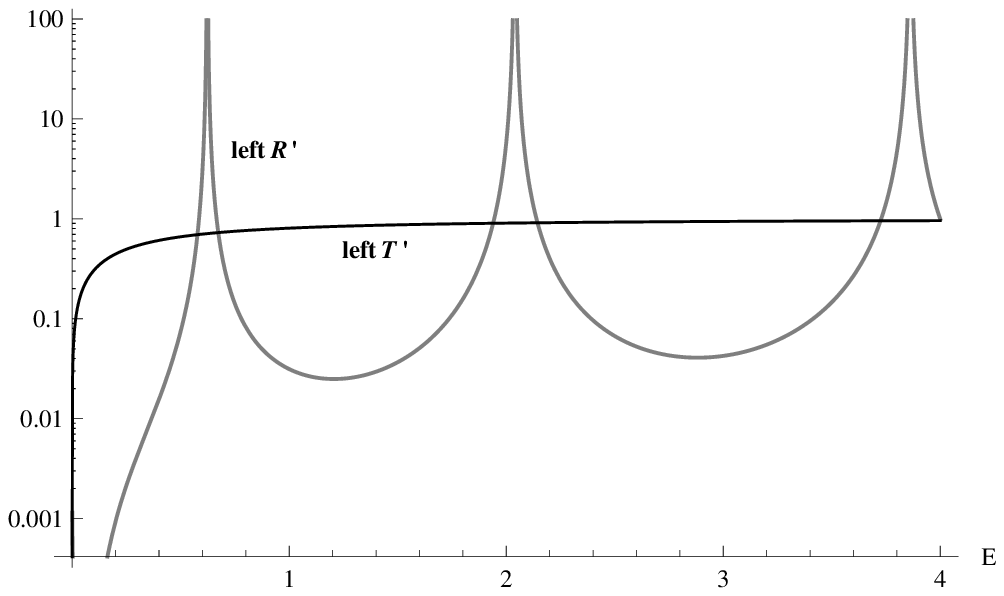} \\

{\it Fig. 2: Critical couplings for left incident waves against incident energies are shown 
for the potential $\tilde{V}(\bar{x})$ in Fig. 2(a) and its time reversal partner 
$\tilde{V}^*(\bar{x})$ in Fig. 2(b). 
In Fig. 2(a) both $R_l$ and $T_l$ vanish at the energies $E_n^l$ with $n_{min}=3$ (for
$V_0 = 1.2, \rho  = 1.8$ and $m = 1$). On the other hand Fig. 2(b) shows divergence of
total scattering coefficient (as $R'_l\rightarrow \infty$) at the same incident energies 
for the time reversed potential. } \\

\subsection{CC for right incident wave}

In this subsection we would like to emphasis that condition for CC depends on the direction
of incidence. In particular we show
that condition of CC for left incidence is different from that of for right incidence for
gain and loss symmetric non-Hermitian WS potential. 
From Eqs. (\ref{t}) and (\ref{rr}) the scattering coefficients $R_r=\mid r_r\mid^2=\mid\-\frac{G1}{G3}\mid^2 $ and $T_r=\mid t_r\mid^2=\mid\frac{1}{G3}\mid^2$   vanish simultaneously when
$2a_2$ is a positive integer ($n'$). 
This leads to CC at the energy, $E_{n'}^r=\frac{\rho  ^{2}}{16m} n'^{2} $, 
which is different from $E_n^l$ at which CC occurs for left incidence. This happens
due to the asymmetry in the left and right asymptotic behavior of the potential. However the 
energy separation between two consecutive CC is independent of direction of incident wave
for same values of $n$ and $n'$ as,
\begin{equation}
\Delta E_{n'}^r\equiv E_{n'+1}^r-E_{n'}^r=\frac{\rho  ^{2}}{16m} (2n'+1)= \Delta E_{n=n'}^l \ .
\label{esr}
\end{equation} 
Following the discussion of the previous subsection it can be shown easily that null scattering due
to $\tilde{V}(\bar{x})$ is same as the super scattering due to $\tilde{V}^*(\bar{x})$ at the same energy even for right incidence. This situation is nicely demonstrated in Fig. 3.

\includegraphics[scale=.8]{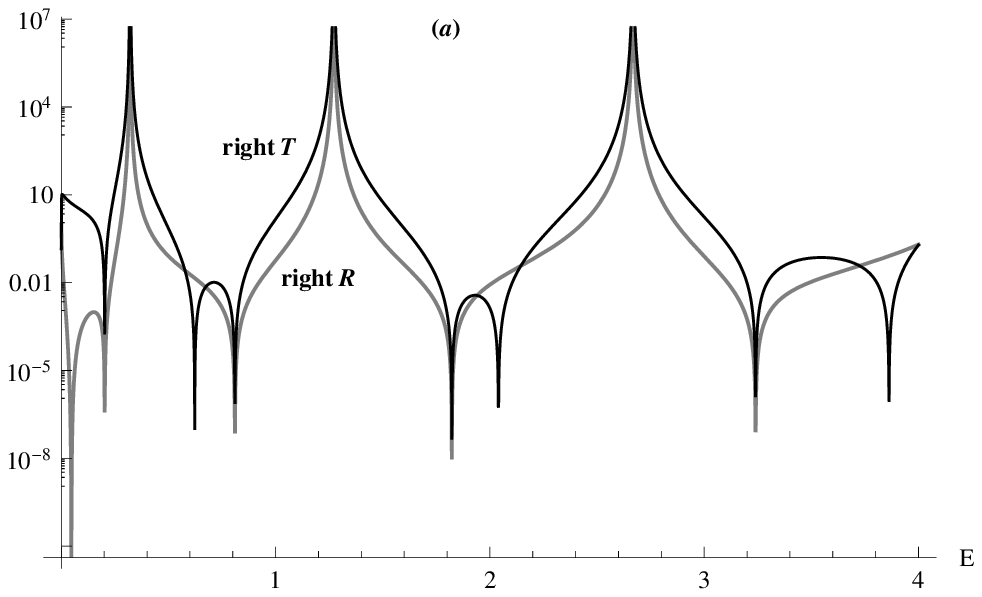} \ \includegraphics[scale=.78]{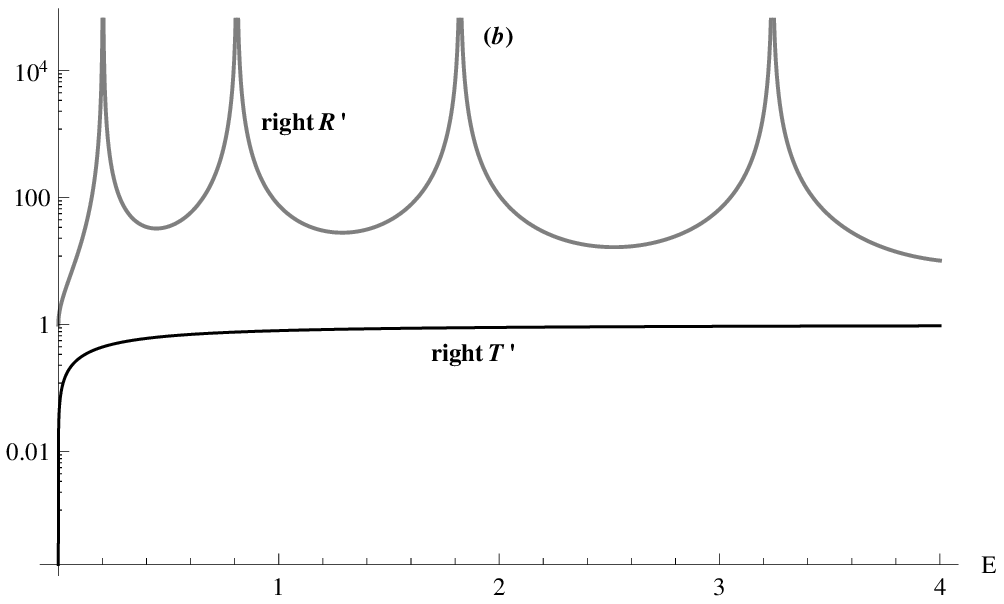}  \\

{\it Fig. 3: Critical couplings for right incidence shown for gain and loss symmetric WS potential.
In Fig. 3(a) both $R_r$ and $T_r$ for $\tilde{V}(\bar{x})$
vanish at the energies 
$E_n^r$ with $n=1,2,3..$ (for
$V_0 = 1.2, \rho  = 1.8$ and $ m = 1$). On the other hand Fig. 3(b) shows the divergences
of $R'_r$  at the same incident energies
for $\tilde{V}^*(\bar{x})$. } \\

Thus the gain and loss symmetric WS potential potential can work as 
a critical coupler for both left and right incidences but for different frequencies. 
The waves with these frequencies when incident from left or right on the time reversed gain and loss symmetric WS potential result in super 
reflectivity 
($R'_l\rightarrow \infty$ or $R'_r\rightarrow \infty$, with finite $T'$).

\section{Critical coupling for ranges of incident energies}

In this section we find critical coupling for different energy ranges. To 
demonstrate this we would like to consider time reversed potential $\tilde{V}^*(\bar{x})$ for which 
the transmission and reflection amplitudes are written in Eqs. (\ref
{rl2})-(\ref{rr2}). For this time reversed case left handed transmission amplitude neither vanishes
nor diverges as $a_2, a_3$ are real positive numbers. However we can adjust the values of
the parameter $\rho $ and $V_0$ in such a manner that $T'\equiv \mid t'_{l,r}\mid^2$ becomes negligibly small over
an interval of energy. This is achieved by considering $\rho $ very small such that
$a_2$ and $ a_3$ are very high and by adjusting $V_0$ in such a manner that $a_3\gg  a_2$. In that case
the dominating term $\Gamma(2a_3)$ in the denumerator of the transmission amplitude will be 
very high leading to very small or negligible transmission. At the same time left handed
reflection amplitude also becomes negligible for a certain interval of energy due to 
the presence of $\Gamma(2a_3)$ term in denumerator. Simultaneously $R'_l\equiv \mid r'_{l}\mid^2$ also vanishes for discrete 
positive energies at which $a_2-a_3=-n$ condition is satisfied. 
But this interval is interrupted by certain singularities when $2a_3=N$ occur. 
Both n and N are positive integers. 
The positive discrete energies at which $R'_l$ diverges are written as,
\begin{equation}
E_{ss}^N=\frac{N^2\rho ^2}{16 m}-V_0, \ \ \ \mbox{for} \ \ N>\frac{4}{\rho }\sqrt{mV_0} \ .
\label{ess}
\end{equation}
Energy gap between two such successive SS is,
\begin{equation}
\bigtriangleup E_{ss} =E_{ss}^{N+1}-E_{ss}^{N}=(2N+1)\frac{\rho ^2}{16m} \ .
\label{essd}
\end{equation} 
Therefore between two consecutive spectral 
singularities we have a certain energy range where $R'_l\approx 0, T'_l\approx 0$. This range is linearly 
increasing with the integer values of N, and also can be controlled by the parameters $\rho $ 
and $V_0$. In such intervals of energy $R'_l$ becomes exactly zero at certain discrete positive energies
(for $a_2-a_3=-n$) which are 
calculated as,
\begin{eqnarray}
E- \sqrt{E^2+EV_{0}}&=&\frac{1}{2} \Big( \frac{n \rho }{2 \sqrt{m}}\Big)^{2} -\frac{V_{0}}{2}\equiv p_n \ , \nonumber \\
\mbox{i.e.} \ \ E_n&=&\frac{p_n^{2}}{V_{0}+2p_n} \ .
\label{er0}
\end{eqnarray}
These discrete energies $E_n$ is real and positive for 
$n>\sqrt{\frac{4mV_0}{\rho ^2}}$. The separation between two such consecutive zeros of $R'_l$ is computed as,
\begin{equation}
\label{sep0}
E_{n+1}-E_{n}=(2n+1) \Big(\frac{\rho ^{2}}{18 m}- \frac{mV_{0}^{2}}{n^{2} (n+1)^{2}\rho ^2}  \Big) \ .
\end{equation}
Thus in case of $\tilde{V}^*(\bar{x})$ we obtain different range of CC separated by SS
for left incidence. On the other 
hand for $\tilde{V}(\bar{x})$ we obtain ranges of SS separated by CC at the same energy values. All
these features of WS potential are well illustrated in Fig. 4. However we would like to 
point out no such ranges of energy exist for $\tilde{V}^*(\bar{x})$ in the case
of right incidence.

\includegraphics[scale=0.8]{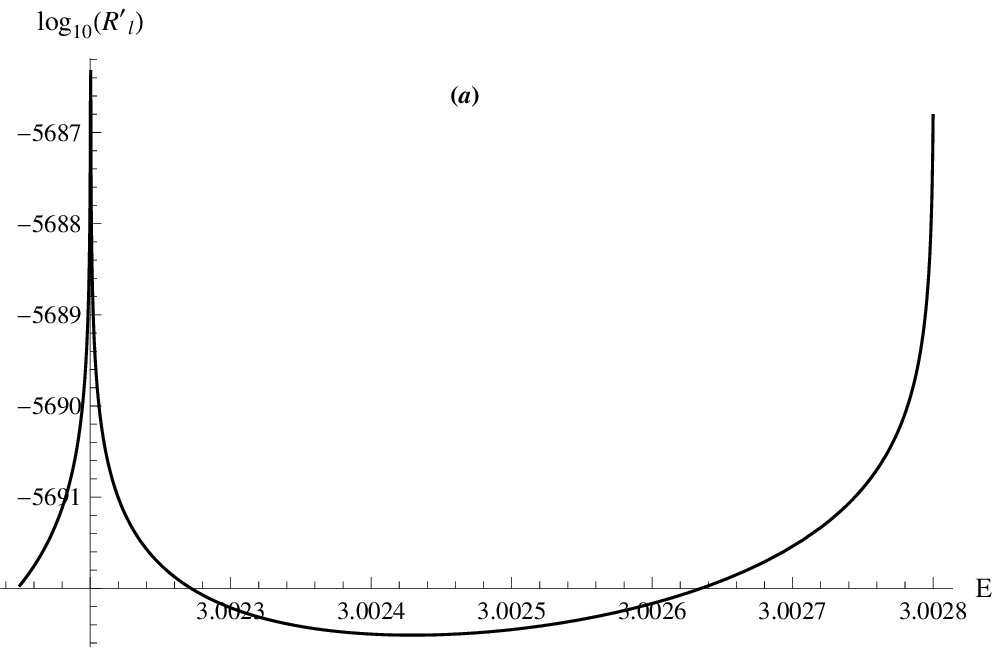} \ \ \ \includegraphics[scale=0.8]{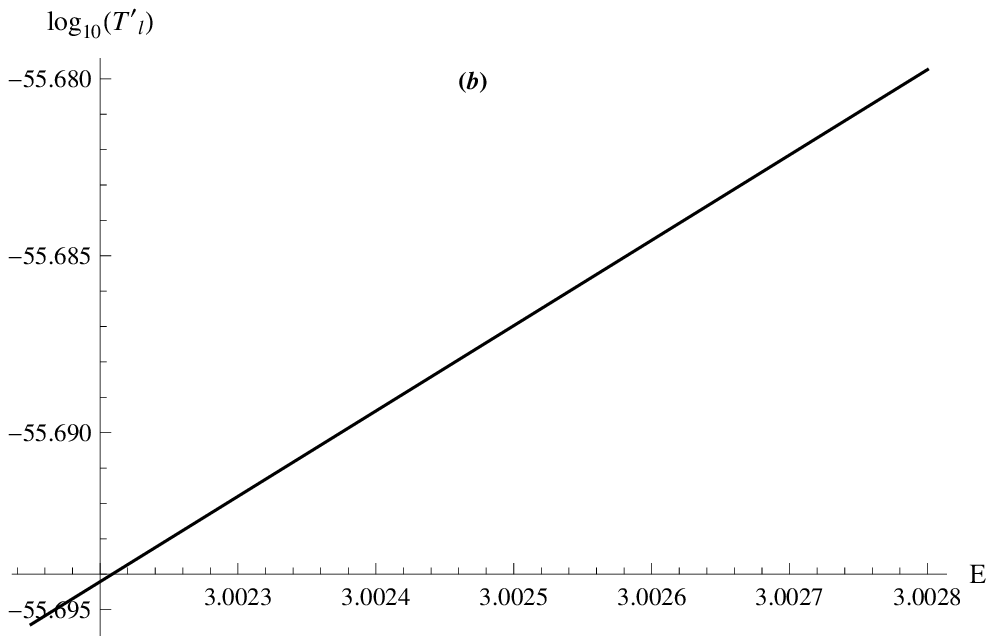} \\

\includegraphics[scale=0.8]{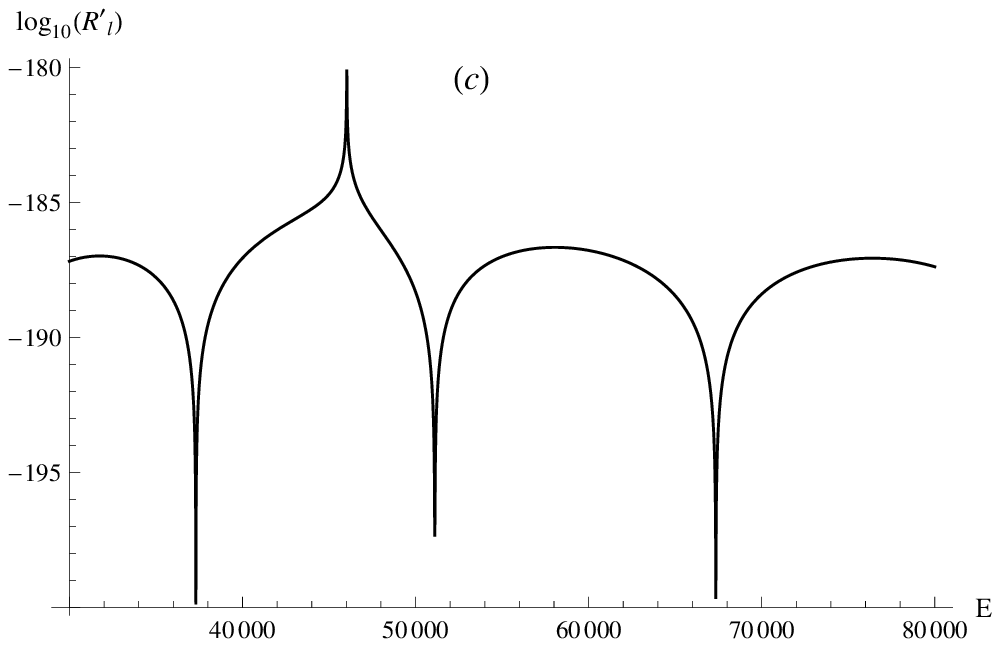} \ \ \ \includegraphics[scale=0.8]{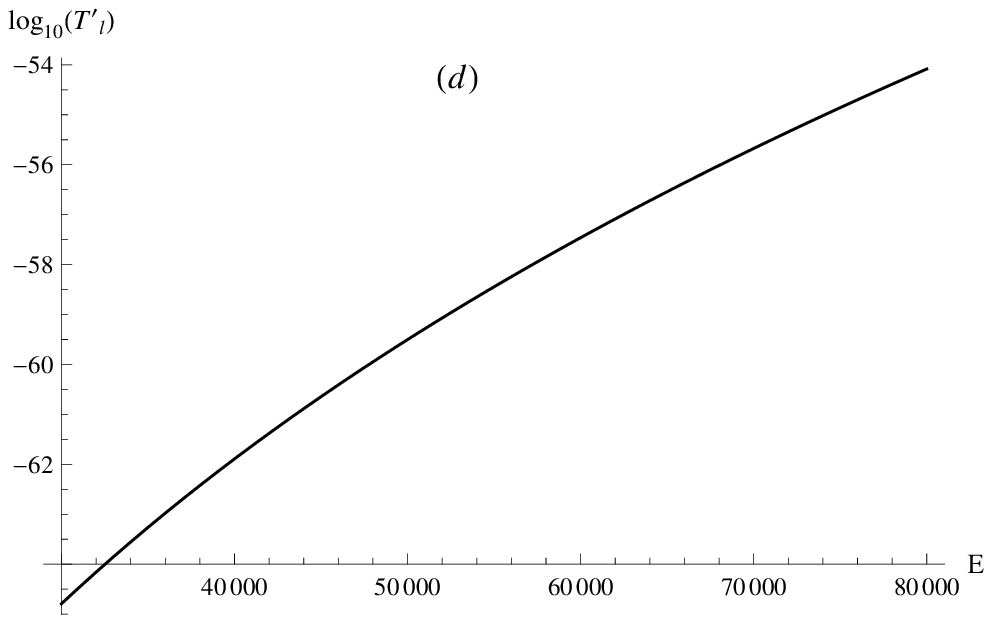} \\

{\it Fig. 4: $log_{10}(R'_l)$ and 
$log_{10}(T'_l)$ are plotted to demonstrate different energy ranges of CC due to the
potential $\tilde{V}^*(\bar{x})$. 
In Figs. 4(a) and (b) one particular range of energy is shown between two successive SS ($\rho =.0006, V_0 = 1, m = 1$) where $R'_l\approx 0$ and  $T'_l\approx  0$. 
Different energy ranges for $R'_l\approx 0$ are shown in Fig. 4(c) (for $\rho =60; V_0 = 5.5\times 10^{6}, m = 1$),  whereas 
Fig. 4(d) shows that $T'_l$ is vanishingly small in those intervals.
Any desired ranges of incident energy for CC can be achieved with the 
potential $\tilde{V}^*(\bar{x})$ by adjusting the parameters in the potential. 
} \\

\section{Perfect absorption of bidirectional waves}
In this section we investigate the absorption through this particular potential distribution when waves are 
coming from both directions. As mentioned in the introduction perfect absorption
for bidirectional waves will occur if $\Big| t_lt_r -r_lr_r=0\Big|$. From Eqs. (\ref{rl})-(\ref{rr}) 
we obtain, 
\begin{equation}
\frac{1}{(G3)^2}\Big[\frac{k_1}{k_2}+G1 G4\Big]=0 \ .
\label{g2g3}
\end{equation} 
Using the properties of $\Gamma$ function we find the relation $G4 G1+\frac{k_{1}}{k_{2}}= 
G2 G3$. Thus perfect absorption occurs for the bidirectional waves if 
$\frac{G2}{G3}=0 \ (G3\not=0)$. Since $a_2, a_3$ are real positive numbers, CPA will occur for 
the potential $\tilde{V}(\bar{x})$ when $G3$ is infinity ($G2$ can not be zero for $\tilde{V}(\bar{x})$). From
Eq. (\ref{gg}) we see that this situation occurs in two possible ways either 
$2a_2=n_1+1$ or $2a_3=n_2$. Thus we obtain CPA for $\tilde{V}(\bar{x})$ at two different energies
\begin{equation}
E_{n_1}=\frac{(n_1+1)^2\rho ^2}{16 m} \ \ \mbox{and} \ \ E_{n_2}=\frac{n_2^2\rho ^2}{16 m}-V_0 \ , 
\end{equation}
where $n_1, n_2$ are positive integers. At energies $E_{n_1}$, $T=0, R_l=0$ and at energies 
$E_{n_2}$, $T=0, R_r=0$ [as
shown in Fig. 5].
Alternatively for the time reversed case  $a_2$ and
$a_3$ changes sign and CPA can only occur for $G2=0$ ($G3$ is always finite in the case of $\tilde{V}^*(\bar{x})$).
The discrete positive energies for CPA are calculated as,
\begin{eqnarray}
\label{cpae}
E+\sqrt{E^2+EV_0}&=&\frac{M^2\rho ^2}{8}-\frac{V_0}{2}\equiv q_M \ , \nonumber \\
 \ \mbox{i.e.} \ \ E&=&E^*_M=\frac{q_M^2}{V_0+2q_M} \ (\mbox{where M is positive integer}).  
\end{eqnarray}
To ensure real positive energy $M$ must be greater than $\sqrt{\frac{4mV_0}{\rho ^2}}$. This depicts that at least one such discrete 
positive energy for CPA
will exist for the potential $\tilde{V}^*(\bar{x})$ if $\rho^2>\frac{|V_0|}{4}$. At these energies left incident
and right incident waves interfere destructively and absorbed by the potential completely. Waves with energies $E_M^*$ when incident on the potential $\tilde{V}(\bar{x})$ produce 
lasing behavior where all the scattering amplitudes diverge (as $a_2+a_3$ is equal to a positive integer). 
Similarly for $\tilde{V}^*(\bar{x})$, $R'_l\rightarrow \infty$ and $R'_r\rightarrow \infty$ at energies $E_{n_1}$ and 
$E_{n_2}$ respectively where CPA occur for $\tilde{V}(\bar{x})$. Thus spectral singularity of $\tilde{V}^*(\bar{x})$ and
CPA of $\tilde{V}(\bar{x})$ occur at the same energy points. \\ \\

\includegraphics[scale=0.79]{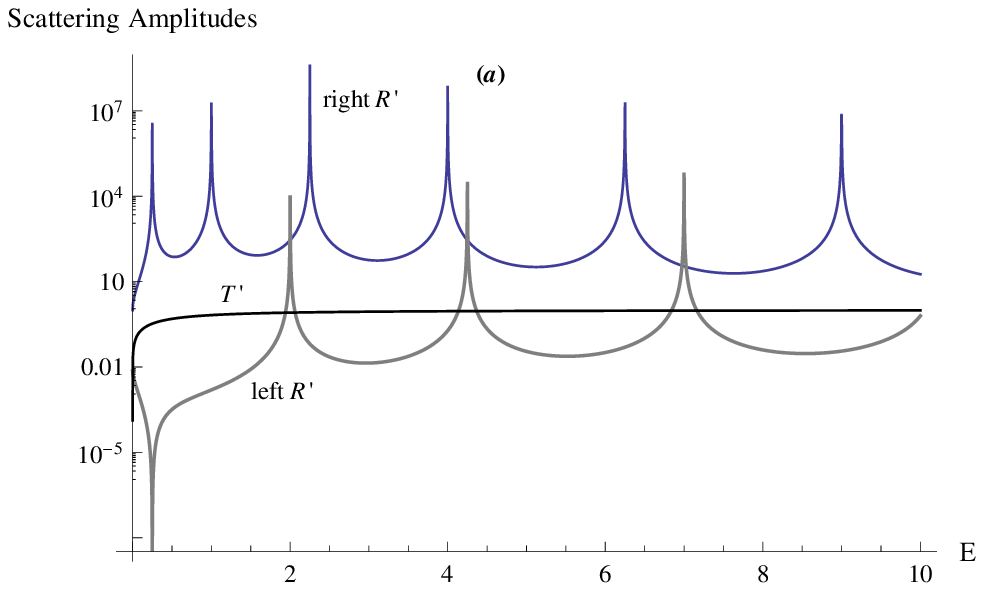} \ \ \ \includegraphics[scale=0.79]{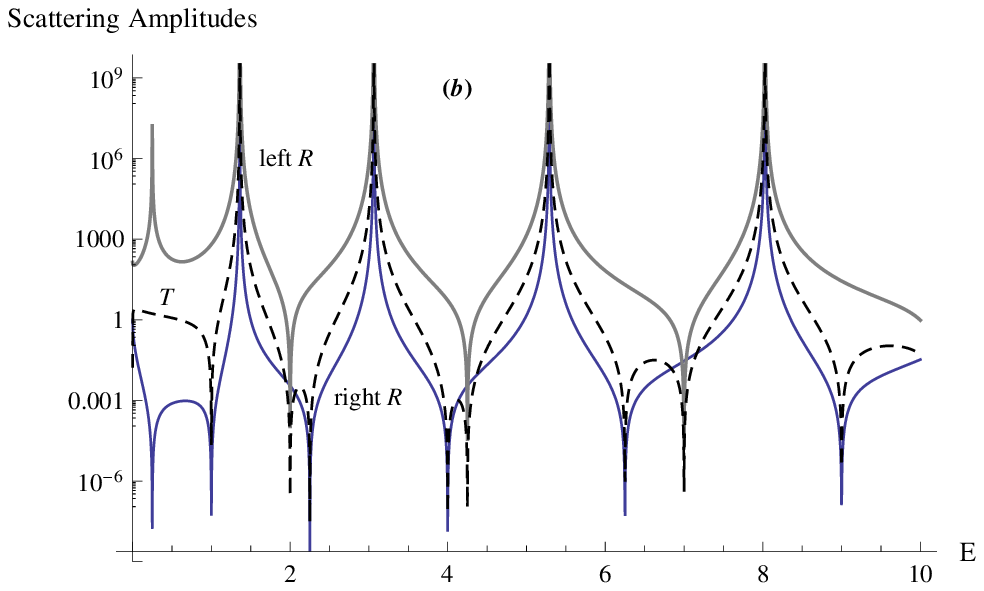} \\

\includegraphics[scale=0.76]{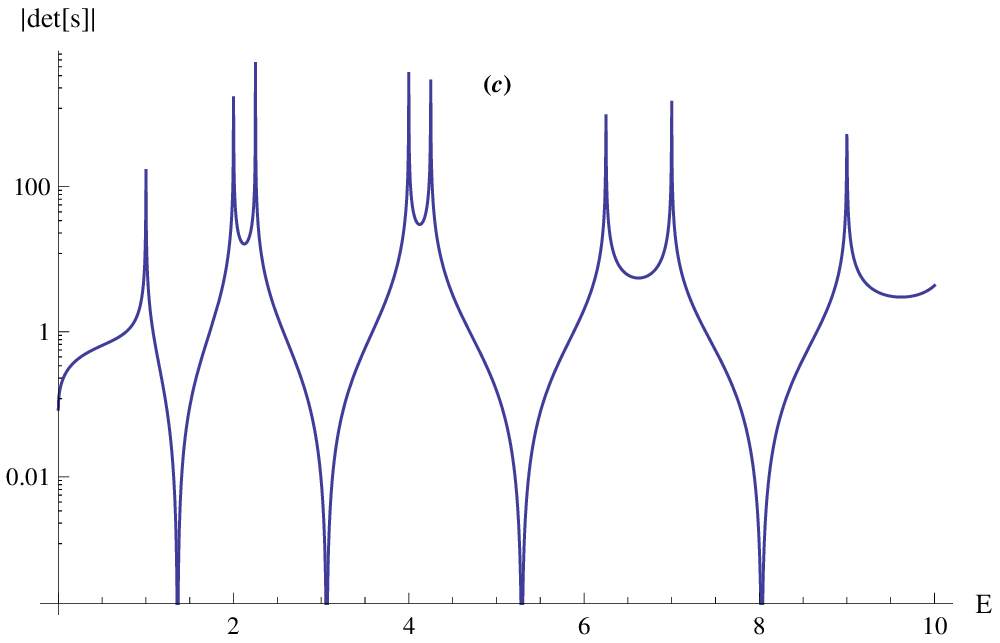} \ \ \ \includegraphics[scale=0.76]{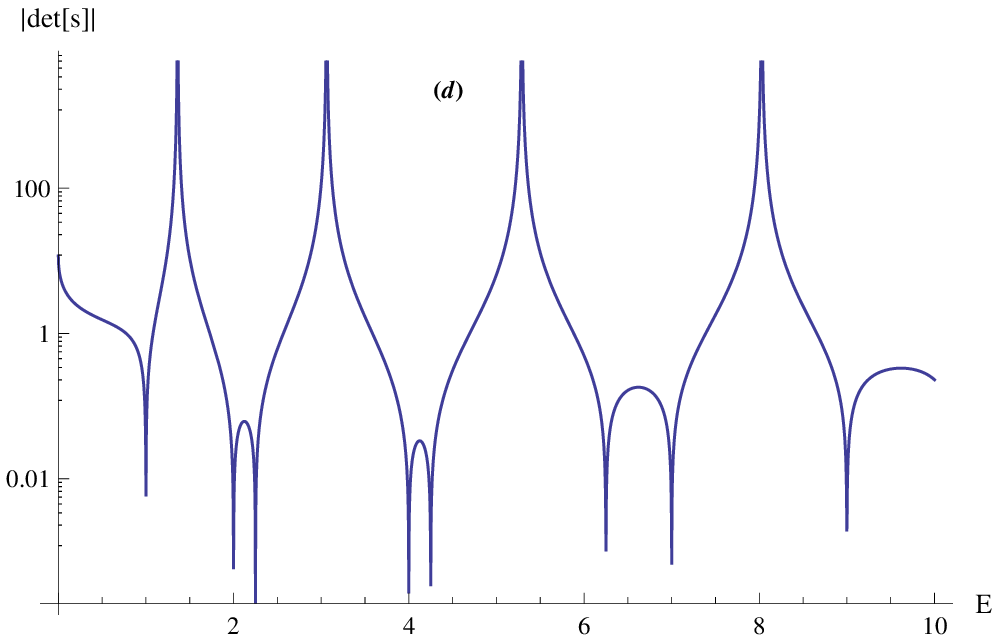} \\

{\it Fig. 5: Figs (a) and (c) are for the potentials 
$\tilde{V}^*(\bar{x})$ whereas (b) and (d) are for $\tilde{V}(\bar{x})$. Behavior of R and T are shown in
(a) and (b) for $V_0 = 2, \rho  = 2, m = 1$. On the other hand CPA and its time reversed situation are shown in (c) and (d).  } \\

To obtain the ranges of energy for CPA for the potential $\tilde{V}^*(\bar{x})$ we choose a 
small $\rho $ such that $a_2, a_3 $ are large even for lower energies.
Then we adjust $V_0$ in such a manner that $a_3$ is large 
enough compare to $a_2$. In such situation we obtain a range of energy for which $t'_{r,l}\approx 0$ and $r'_l\approx 
0$ and hence $\Big|det[S]\Big|\approx  0$, leading to CPA. This range is interrupted by the singular points of $r'_l$ and $r'_r$
which occurs due to the presence of $\Gamma(-2 a_3)$ and $\Gamma(1-2 a_2)$ in the numerator of $r'_l$ and $r'_r$ respectively. Thus range of CPA is separated by these SS as shown in Fig. 6.
The $n^{th}$ spectral singular point for $r'_l$ is at the positive discrete energies $E_n=\frac{n^2\rho ^2}{16m}-V_0$ for which $2a_3=n$, with $n_{min}=Int(\sqrt{\frac{16m V_0}{\rho ^2}})+1$. The energy interval for any two consecutive singularities of $r'_l$ is, 
$$\bigtriangleup E_{ss}^{rl}=E_{n+1}-E_{n}=\frac{\rho ^2}{16m}(1+2n).$$ 
On the other hand  
$r'_r$ has singularities for $2a_2=n'$ (i.e. at $E_n'=\frac{n'^2\rho ^2}{16m}$, where $n'$ is another
positive integer) which occur at the energy intervals of, $$\bigtriangleup E_{ss}^{rr}=E_{n'+1}-E_n'=\frac{\rho ^2}{16m}(1+2n').$$ 
Both the energy intervals occurs periodically and $\bigtriangleup E_{ss}^{rl}>\bigtriangleup E_{ss}^{rr}$ as $n>n'$.  
The energy span of these ranges increase with increasing $n'$ for a fixed 
$\rho $. Fig. 6 shows the ranges of energies of perfect absorption for the time reversed potential with
different parametric regimes. \\

\includegraphics[scale=0.76]{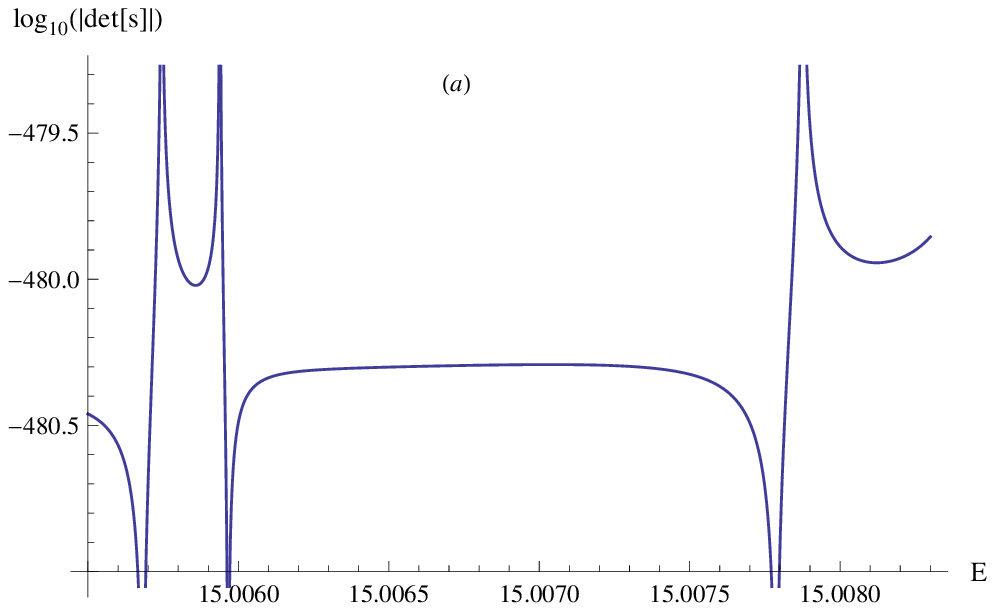} \ \ \ \includegraphics[scale=0.76]{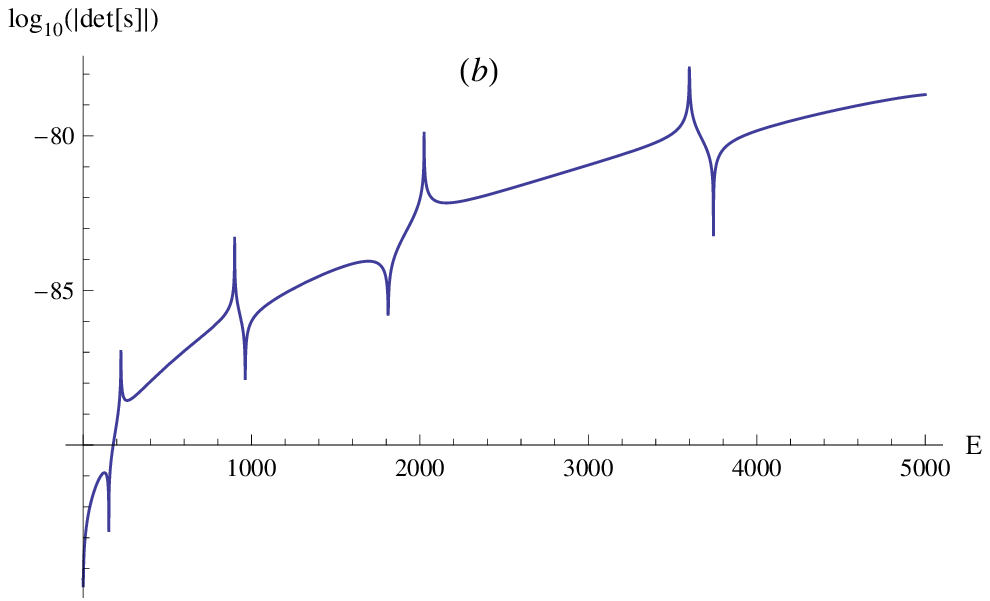}

{\it Fig. 6: Energy ranges for CPA  
for the potential $\tilde{V}^*(\bar{x}))$ are demonstrated. Fig. 6(a) shows a range of energy
with $\rho  = .001, V_0 = 15, m=1$. In Fig. 6(b) we have shown periodic ranges of 
energies separated by spectral singular points for a different parametric regime ($\rho  = 
60; V_0 = 5.5\times 10^{6}, m=1)$. } \\

\section{Conclusions and discussion}
It worths studying CC and CPA and its properties for interaction 
of particle waves with various non-Hermitian models to search for new features and
possibilities for perfect absorption.
In this work we have shown that for a particular gain and loss 
symmetric 
non-Hermitian optical potential (WS potential) it is possible to achieve CC and CPA for a 
range of frequencies due to quantum scattering. 
We have obtained that the conditions of CC depend on the direction of incident waves and no 
range exists for right incident case for this non-Hermitian potential. More interestingly by adjusting the parameters in the 
potential we can have these total
absorptions in any desired ranges of frequencies. For a PT-symmetric non-Hermitian optical 
potential we have derived the analytical conditions of null scattering (CC and CPA) to occur. 
The energy 
points at which the null scattering occurs are shown to be the SS points for the time 
reversed system. An estimation of the values of parameters are provided in 
in Table. 1. The CC and CPA energies and energy ranges increases with increasing shape parameter, i.e. decreasing diffuseness of the potential. As this specific 
potential plays an important role in describing interactions of nucleon with heavy nucleus 
\cite{wsu1,wsu2}, our
theoretical demonstration will open the possibility of future work on absorption of interacting
microscopic particles of different masses and energies. \\

\begin{table}
\caption{Energy and energy ranges for CC and CPA for different values of 
parameters (with incident particle mass $m=1$, in atomic unit) of 
non-Hermitian PT-symmetric WS potential.}
\centering
\begin{tabular}{c c c c c}
\hline\hline 
$V_0$  & \ \ $a=1/\rho $  & \ \ $r_0=\pi .a$  &  Incident particle energies  & 
Corresponding \\[0.5ex]
 & \ \ (in nm) & \ \ (in nm) & \& energy ranges for CC and CPA  &  Figs.\\[0.5ex]
\hline\hline
32.65 eV &	.029 &	.093 &	$E_3^l=16.94 $ eV, $E_4^l=55.50$ eV, & Fig. 2 \\
& & & \ \ $E_5^l=105.07$ eV.
\\[0.2ex]
\hline
32.65 eV &	.029 &	.093 &	$E_1^r=5.51 $ eV, $E_2^r=22.04$ eV, & Fig. 3 \\
& & & $E_3^r=49.58$ eV. 	
\\[0.2ex]
\hline
27.2 eV & 88.3 & 277.5	& Range of CC $\cong$ .0006 eV; & Fig. 4 (a,b)\\
& & & (E=81.6791 eV to 81.6954 eV) \\[0.3ex]
150 MeV & $.0009$ & $.003$ & Range of CC $\cong$ .46 MeV. & Fig. 4 (c,d)\\
& & & (E=1.37 MeV to 1.83 MeV) \\
[0.2ex]
\hline
54.41 eV &	.02 & .08 &	$E_{n1}=27.2$ eV$(n1=1)$,  & Fig. 5(b,d)\\
& & & $E_{n2}=54.4$ eV $(n2=4)$,\\
& & & $E_{n1}=61.2$ eV$(n1=2)$. \\
[0.2ex]
\hline	
54.41 eV &	.02	& .08 &	$E^*_M=37.03$ eV$(M=3)$, & Fig. 5(a,c)\\
& & & $E^*_M =83.32$ eV$(M=4)$, \\
& & & $E^*_M =143.92$ eV$(M=5)$.	\\
[0.2ex]
\hline	
408.01 eV &	53	& 166.5 & Range of CPA$\cong $.053 eV; & Fig. 6(a)\\
& & & (E=408.096 eV to 408.258 eV) \\[0.3ex]
150 MeV & $.0009$ & $.003$ & Range of CPA$\cong $.042 MeV. & Fig. 6(b)\\
& & & (E=.055 MeV to .097 MeV) \\
[1ex]
\hline
\end{tabular}
\end{table}

\vspace{.5in}

{\bf Acknowledgments}: Two of the authors (A. Ghatak and B. P. Mandal) acknowledge the 
hospitality of TIFR, Mumbai, where a part of revision of this manuscript is done.


\begin{thebibliography}{99}

\bibitem{ben4} C. M. Bender and S. Boettcher {\em Phys.Rev.Lett.} {\bf 80}, 5243 (1998).
\bibitem{mos} A. Mostafazadeh, {\em Int. J. Geom. Meth. Mod. Phys.} {\bf 7}, 1191(2010) and references therein.
\bibitem{ben} C. M. Bender, {\em Rep.Prog. Phys.} {\bf 70}, 947 (2007) and references therein.


\bibitem{opt1} Z. H. Musslimani, K. G. Makris, R. El-Ganainy, and D. N. Christodoulides {\em Phys.Rev. Lett.} {\bf 100}, 030402 (2008).

\bibitem{opt2} C. E. Ruter, K. G. Makris, R. El-Ganainy, D. N. Christodoulides, M. Segev, D. Kip, {\em Nature Physics} {\bf 6} 192, (2010); 

\bibitem{opt3}  R. El-Ganainy, K. G. Makris, D. N. Christodoulides and Z. H. Musslimani {\em Optics Letters} {\bf 32}, 2632 (2007).

\bibitem{eqv1} A. Guo et al {\em PRL} {\bf 103}, 093902 (2009).

\bibitem{bm} P. Siegl and D. Krejcirik {\em Phys. Rev. D} {\bf 86}, 121702 (2012).

\bibitem{bmm} B. Basu-Mallick, {\em Int. J. of Mod. Phys. B}  {\bf 16}, 1875 (2002); B. Basu-Mallick, T. Bhattacharyya, A. Kundu, and B. P. Mandal {\em Czech. J. Phys } {\bf 54}, 5 (2004).

\bibitem{sca1} B. Basu-Mallick and B.P. Mandal, {\em Phys. Lett. A} {\bf 284}, 231 (2001); B. Basu-Mallick, T. Bhattacharyya  and B. P. Mandal, {\em
 Mod. Phys. Lett. A} {\bf 20 }, 543 (2004).
 
\bibitem{new3} A. Khare and B. P. Mandal, {\em Phys.Lett.} {\bf A272}, 53 (2000).
\bibitem {bsg} B. P. Mandal, S. Gupta, {\em Mod.Phys.Lett. A} {\bf 25}, 1723 (2010).
\bibitem{bpm} B. P. Mandal, {\em Mod. Phys. Lett. A} {\bf 20}, 655(2005).

\bibitem{ent} A. Ghatak and B. P. Mandal, {\em J. Phys. A: Math. Theor.}
{\bf 45}, 355301 (2012).

\bibitem{cal} B. P. Mandal and A. Ghatak, {\em J. Phys. A: Math. Theor.}
 {\bf 45}, 444022 (2012) .

\bibitem{ph} B. P. Mandal, B. K. Mourya, and R. K. Yadav (BHU),
{\em Phys. Lett. A} {\bf 377}, 1043 (2013).

\bibitem{sol} S. Longhi {\em Phys. Rev. B}  {\bf 80}, 165125 (2009).

\bibitem{opap} H. Ramezani, T. Kottos, R. El-Ganainy, D. N. Christodoulides, {\it 
Phys. Rev. A} {\bf 82}, 043803 (2010). 

\bibitem{opapp} Y. D. Chong, Li Ge, and A. D. Stone, {\it Phys. Rev. Lett.} {\bf 
106}, 093902 (2011).


\bibitem{ep0} T. Kato, {\em Perturbation Theroy of Linear Operators}, {\bf Springer}, Berlin, (1966).
\bibitem{ep1} M.V. Berry, {\em Czech. J. Phys.} {\bf 54}, 1039 (2004).
\bibitem{ep2} W. D. Heiss, {\em Phys. Rep.} {\bf 242}, 443 (1994).


\bibitem{ss1} A. Mostafazadeh {\em Phys. Rev. Lett.} {\bf 102}, 220402 (2009).

\bibitem{ss2} Ali Mostafazadeh, Mustafa Sarisaman, {\em Phys. Lett. A} {\bf 375}, 3387 (2011).

\bibitem{aop} A. Ghatak, R. D. Ray Mandal, B. P. Mandal, {\em Annals of Physics} {\bf 336}, 540 (2013).

\bibitem{ss3} A. Ghatak, J. A. Nathan, B. P. Mandal, and Z. Ahmed, {\em J. Phys. A: Math. 
Theor.} {\bf 45},  465305 (2012).

\bibitem{inv2}  S. Longhi, {\em J. Phys. A: Math. Theor.} {\bf 44}, 485302 (2011).

\bibitem{inv1} A. Mostafazadeh, {\em Phys. Rev. A} {\bf 87}, 012103 (2013).

\bibitem{inv3} Z. Ahmed, C. M. Bender, M. V. Berry, {\em J.Phys.A} {\bf 38}, L627 (2005).

\bibitem{resc} L. Deak, T. Fulop, {\em Annals of Phys.} {\bf 327}, 1050 (2012).

\bibitem{cpa00} C. F. Gmachl, {\em NATURE} {\bf 467}, 37 (2010).

\bibitem{cpa01} S. Longhi {\em Physics 3} {\bf 61} (2010).

\bibitem{cpa011} W. Wan, Y. Chong, L. Ge, H. Noh, A. D. Stone, H. Cao, {\em SCIENCE} {\bf 331}, 889 (2011).

\bibitem{cpa02} N. Liu, M. Mesch, T. Weiss, M. Hentschel, and H. Giessen, {\em Nano Lett.} {\bf 10}, 2342 (2010).

\bibitem{cpa1} H. Noh, Y. Chong, A. Douglas Stone, and Hui Cao, {\em Phys. Rev. Lett.} {\bf 108}, 6805 (2011).

\bibitem{cpa0} A. Mostafazadeh and M. Sarisaman, {\em physics.optics} {\bf 24} (2012).

\bibitem{cpa2} S. Longhi, {\em Phys. Rev. A} {\bf 83}, 055804 (2011).

\bibitem{cpa3} S. Dutta-Gupta, R. Deshmukh, A. Venu Gopal, O. J. F. Martin, and S.
Dutta Gupta, {\em Optics Letters} {\bf 37}, 4452 (2012).

\bibitem{cpa4} N. Liu, M. Mesch, T. Weiss, M. Hentschel, and H. Giessen, {\em Nano Lett.} {\bf 10} 2342 (2010).


\bibitem{cc0} M. Cai, O. Painter, and K. J. Vahala, {\em Phys Review Letter } {\bf 85}, 74 (2000).

\bibitem{cc1} J. R. Tischler, M. S. Bradley, and V. Bulovic, {\em Optics Letters} {\bf 31},
2045 (2006)

\bibitem{cc2} S. Dutta Gupta, {\em Optics Letters} {\bf 32}, 1483 (2007).

\bibitem{cc3} S. Balci, C. Kocabas, and A. Aydinli, {\em Optics Letters} {\bf 36}, 2770 (2011).

\bibitem{cc4} S. Balci, Er. Karademir, C. Kocabas, and A. Aydinli, {\em Optics Letters} {\bf 36}, 3041 (2011).

\bibitem{wsu1}  A. Diaz-Torres, and W. Scheid, {\em Nucl. Phys. A} {\bf 757}, 373 (2005).
\bibitem{wsu2}  J. Y. Guo, and Z. Q. Sheng, {\em Phys. Lett. A} {\bf 338}, 90 (2005).

\bibitem{wsbs} R. Id Betan and W. Nazarewicz, {\em J. Phys.: Conf. Ser.} {\bf 436}, 012061 (2012); 

\bibitem{wsdf} W. D. Myers and H. Von Groote, {\em Physics Letter B} {\bf 61}, (1976).

\bibitem{wsds} Yu-Jun Mo, Sheng-Qin Feng and Ya-Fei Shi, {\em Physical Review C} {\bf 88}, 024901(2013). 

\bibitem{ws2} A. Arda, O. Aydogdu, and R. Sever, {\em J. Phs. A: Math Theor.} {\bf 43}, 425204(2010).

\bibitem{ims} G. Levai, M. Znojil, {\em Mod. Phys. Letters A} {\bf 16}, (2001).

\bibitem{imsh} Levai G, Sinha A and Roy P {\em J. Phys. A: Math. Gen.} {\bf 36}, 7611 (2003).
















\end{thebibliography}
\end{document}